\documentclass[review]{elsarticle}

\usepackage{hyperref}

\journal{Physica A: Statistical Mechanics and its Applications}

\begin{document} 

\begin{frontmatter}

\title{Fermion-like behavior of elements in their spatial distribution around points of interest}

\author{Josu Martinez-Perdiguero\fnref{myfootnote}}
\fntext[myfootnote]{Email address: jesus.martinez@ehu.eus}

\address{Department of Condensed Matter Physics, Faculty of Science and Technology, University of the Basque Country UPV/EHU, Barrio Sarriena s/n, 48940 Leioa, Spain}

\begin{abstract}

In this paper we analyze the spatial distribution of elements around points of interest. Based on a spatial exclusion principle we model the system by means of a Fermi-Dirac distribution defined by two easily interpretable parameters. By means of image analysis, two real cases are studied and compared to the theory: people in an open-air concert and cars in a mall parking-lot. We show that they closely obey the proposed fermion-like statistics.

\end{abstract}


\end{frontmatter}

\section{Introduction}

In the last years, the application of the theory of physics to the description of social systems has given rise to the field nowadays known as sociophysics \cite{castellano_statistical_2009, galam_sociophysics:_2012}. Statistical physics is the main framework in which sociophysics is based on due to two main reasons: On the one hand social systems usually involve very large numbers of elements (whole populations). On the other hand, there is a high degree of randomness due to the `unpredictable' behavior of each element/individual and of their interactions which stem from their complex decision-making processes and many related factors.

Economic phenomena are very well suited to this treatment and the econophysics field stands on its own as a very active field \cite{de_area_leao_pereira_econophysics:_2017}. Opinion/voter models or cultural dynamics are other widespread examples collecting successes of the application of the same framework \cite{castellano_statistical_2009}. In this work we study yet another area, the so-called crowd behavior science, in which the focus is usually on the collective motion of the system elements be them people (crowd behavior) \cite{karamouzas_universal_2014}, vehicles (traffic)\cite{schadschneider_traffic_2002}, birds (flocking), etc., and their emergent phenomena. We concentrate on situations where the elements are drawn towards a singular spatial position, which we will call the point of interest (POI),  by some `force'. This attractive force could be, although not exclusively, of physical, chemical (e.g., pH gradient) or of an abstract social nature (such as could be the forces of necessity, curiosity, passion, etc). Each element (or agents or particles) under one of these forces will, taking into account the restrictions, minimize their distance to the given POI. For example, in the case of people attending an arts performance (e.g., concerts, a mime in the street, theater or circus) they will try to satiate the feeling by being as close to the stage as possible. Clearly, if the density is high enough, the crowding of elements around these POIs creates a packing problem due to the excluded volume, i.e., one element excludes another one from the effective volume they occupy.

Without attempting to describe the dynamics leading to equilibrium in the statistical sense, in this paper we study the stationary distribution of elements around POIs starting from the hypothesis that they behave obeying an exclusion principle: two elements cannot share the same spatial position/state. This plausible premise lead us to treat them as fermions following the corresponding Fermi-Dirac statistics. The results are successfully applied to two familiar systems: cars in mall parking lot and the audience in a open-air concert.

\section{Theoretical context}
We will model these systems with a physical analogy in which the attraction towards the POI is due to a force $\mathbf{F}(\mathbf{r})$. A first reasonable simplification will be made here assuming that $\mathbf{F}$ only depends on $\mathbf{r}$ through its modulus $r$, i.e., we will assume spherical/radial symmetry. This force stems from a potential field $V(r)$ which, due to attractive nature of the POI, will be a monotonically increasing function of $r$.

We will call the energy $E_i$ the `energy' element $i$ has when they are around a POI. $E_i$ can have several contributions. First, a dependency of $E_i$ on $r_i$ through $V(r_i)$ is expected. Moreover, some other terms accounting for `local fields' at $\mathbf{r}_i$ (e.g., density of agents in the vicinity of $\mathbf{r}_i$) or inner degrees of freedom of element $i$ could be justified. We will adopt the simplest description by restricting ourselves to the case of indistinguishable elements by means of neglecting all terms but the first mentioned so that elements just passively respond to the field.

Following the widespread notation in statistical physics, we define $G(r)$ as the number of possible spatial states at a distance $r$ or less from the POI. Assuming that a continuous treatment is possible, we differentiate to get $dG(r)= g(r)dr$ as the number of states within an interval $dr$ around $r$, where $g(r)$ is the so-called density of states. Analogously, $n(r)$ will be the number of occupied states at $r$ per radius interval. Evidently, the total number of elements will be 
\begin{equation}
N=\int\limits_{\mathrm{all\:space}} n(r)dr
\label{N}
\end{equation}

As mentioned in the introduction, the physical presence of a particle in a given positional state will prevent other (indistinguishable) particles from occupying it so we hypothesize that, in equilibrium, a Fermi-Dirac distribution for the particle positions has to be satisfied:

\begin{equation}
f(r)=\frac{n(r)}{g(r)}=\frac{1}{\exp\lbrace \frac{E(r)-\mu(T)}{kT} \rbrace + 1}
\label{FD}
\end{equation}

\noindent where $kT$ and $\mu(T)$ can be respectively identified with the temperature (up to a constant $k$) and chemical potential in thermodynamic systems  and $f(r)$ gives the so-called occupation number. 

\section{Example I: Open-air music concert}

As a first example we analyze the case of a live-music open-air concert where the center of the stage is considered the POI for the attending crowd. We will consider that the stage faces the public and that the audience area extends radially from a distance $r_0$ from the stage center with an opening angle $\alpha$ (see Figure 1).

Since we are carrying a quantum mechanical treatment, it is interesting to note that we are considering infinite potential walls to delimit the area to analyze. This tries to take into account that people, even when considered as quantum mechanical particles, are not able to go through fences or walls. Mathematically, we can impose a potential such that, in polar coordinates, $V(\mathbf{r})=\infty$ when $\theta \notin [0,\alpha]$ or $r\leq r_0$ resulting in zero probability to find people beyond the fences ($n(\mathbf{r})=0$ in this region).

\begin{figure}
\begin{center}
\includegraphics[scale=0.4]{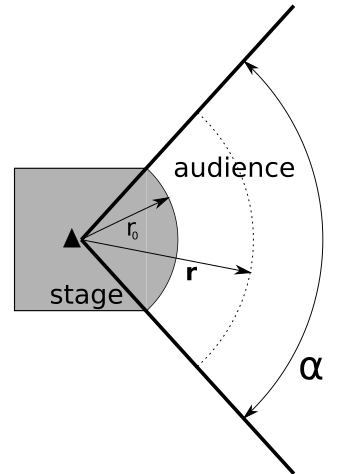}
\caption{Scheme of the proposed geometry for an open-air concert where the audience is located in the area comprised by the angle $\alpha$ starting from a minimum distance $r_0$.}
\label{concsymm}
\end{center}
\end{figure}

We will call $\rho_m$ the maximum standing crowd density whose inverse is the minimum area occupied by a standing person assuming efficient packing, i.e., the so-called mosh-pit density \cite{still_introduction_2014}. We can then calculate $G(r)$ as the product of $\rho_m$ times the area up to $r$:
\begin{equation}
G(r)=\rho_m\int_0^\alpha \mathrm{d\theta} \int_{r_0}^{r}r\mathrm{d}r=\rho_m\alpha\frac{r^2-r_0^2}{2}
\end{equation}
and from here the density of states
\begin{equation}
g(r)=\rho_m\alpha r
\label{gr}
\end{equation}

At this point we can work out explicitly Equation \ref{N} in the low temperature limit $T=0$ when $f(r)$ (Equation \ref{FD}) is non-null and equal to 1 only up to what we will call, following an analogy with non-interacting fermion systems, the Fermi radius $r=r_F$ at which $E(r_F)=\mu(0)$:
\begin{equation}
N=\int_{r_0}^{r_F}g(r)\mathrm{d}r=\rho_m\alpha\frac{{r_F}^2-{r_0}^2}{2}
\label{NrF}
\end{equation}

To go on we need an expression for $E(r)$ giving the `strength' of the POI. We propose here a simple expression with a suitable dependency: $E(r)=A\cdot r$, where the constant $A$ is set to 1 without loss of generality. We have then that $\mu(T=0)=r_F$. As temperature increases from absolute 0, the distribution will change mainly around $r_F$ and in a first approximation (in the low temperature limit) we can treat $\mu(T)$, i.e., the energy level with a 50\% probability of occupation, as a constant equal to its value at $T=0$. A more rigourus approach can be made using the Sommerfeld temperature expansion of $\mu(T)$ \cite{marder_michael_p_condensed_2010} but, in this case, it is not necessary because, as it will be shown later, temperatures are indeed low.

Equation \ref{FD} can be now rewritten as

\begin{equation}
f(r)=\frac{n(r)}{g(r)}=\frac{1}{\exp\lbrace \frac{r-r_F}{T} \rbrace + 1}
\label{FD2}
\end{equation}

\noindent where $k$ has also been set to 1 and with units so that $T$ is a length.

\begin{figure}
\begin{center}
\includegraphics[width=\linewidth]{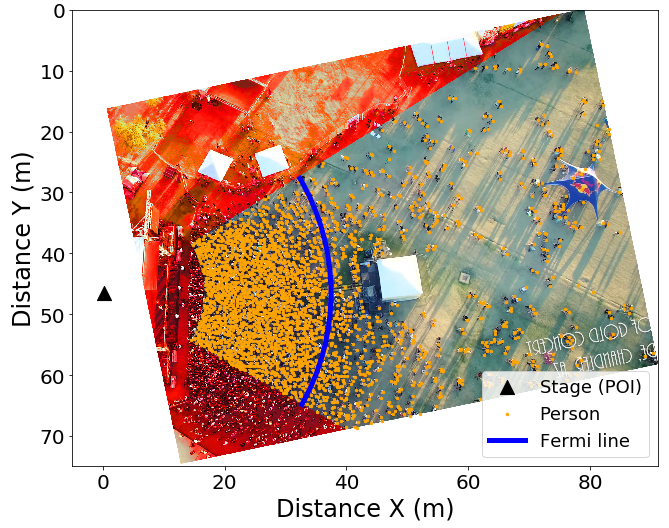}
\caption{Processed aerial photography of an open-air concert \citep{concimage}. For the sake of clarity, the image has been horizontally aligned and the area excluded from the analysis shaded in red. People positions are marked with orange circles and the Fermi line is depicted in blue. The analyzed area has been selected such that the symmetry is conserved while maximizing the angular spread.}
\label{concertexp}
\end{center}
\end{figure}

Based on the above-constructed model we will study the distribution of the public attending a concert of which the vertically taken aerial image shown in Figure \ref{concertexp} was available \cite{concimage}. To directly apply the obtained expressions the area of the image to be analyzed was selected without breaking its symmetry (compare Figures \ref{concsymm} and \ref{concertexp}). The image was processed using the ImageJ image processing and analysis software \cite{schneider_nih_2012}.

Giving the constraints of the photograph, the employed opening angle was 61$^{\circ}$ and distances below $r_0=16.2$ m were not taken into account. In total, 2,721 people were present in the analyzed section.

\begin{figure}
\begin{center}
\includegraphics[width=0.9\linewidth]{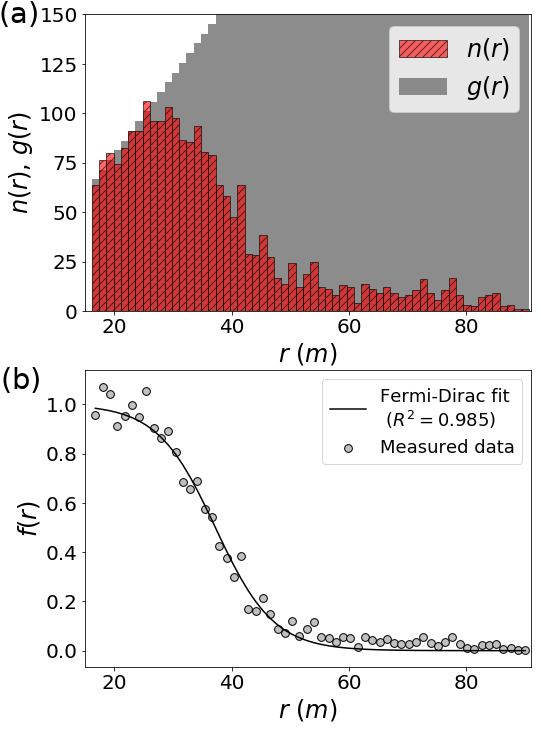}
\caption{(a) Plot of the number of people  $n(r)$ and of the possible states $g(r)$ per distance interval as a function of distance for Example I. The area of Figure \ref{concertexp} has been divided into 60 radial intervals from $r_0$ to the maximum distance available. (b) Plot of the distribution function $f(r)=n(r)/g(r)$ vs. $r$. The dots represent measured data and the solid line a non-linear regression of the data with Equation \ref{FD2}. The best fit parameters are $r_F=37.3\pm 0.3$ m and $T=4.9\pm 0.2$ m.}
\label{conc_ab}
\end{center}
\end{figure}

Figure \ref{conc_ab}a shows a histogram plot of the values of $n(r)$ obtained from the measured data. Since near $r_0$ all positional states are occupied, we have that $f(r\sim r_0)\approx 1$ and the mosh-pit density can be estimated as $\rho_m=\frac{1}{\alpha}\frac{\mathrm{d}n(r)}{\mathrm{d}r}|_{r_0}=3.7\mathrm \pm 0.7\mathrm{\:people/m}$. Using this value in Equation \ref{gr}, $g(r)$ is also calculated and represented in Figure \ref{conc_ab}a.

The distribution function $f(r)$ can be now computed and it is shown in Figure \ref{conc_ab}b. As it can be seen the occupation near the stage is close to unity and at around $r=30$ m it starts declining. It is interesting now to perform a regression with Equation \ref{FD2}, which is shown as the solid line. As it can be seen the agreement is quite good as it closely tracks the experimental data. The parameters of the non-linear fit are $r_F=37.3\pm 0.3$ m and $T=4.9\pm 0.2$ m.

\section{Example II: Cars in a parking lot}
We now work out a second example dealing with the spatial distribution of cars in the mall parking lot of Figure \cite{parkimage}. The empty and occupied parking spots positions were located in the photograph revealing a total of 923 parking spots in front of the supermarket of which 405 were occupied. In this somewhat simpler case $g(r)$ can be obtained by counting the possible parking spots at certain intervals of $r$. In a similar fashion, $n(r)$ is calculated by counting the occupied spots at these intervals. The measured data is shown in Figure \ref{park_ab}a where 40 radial intervals were used. 

From the data of Figure \ref{park_ab}a, the distribution can be computed and it is shown in Figure \ref{park_ab}b together with a nonlinear fit to the Fermi-Dirac distribution (Equation \ref{FD}). The parameters of the fit are $r_F=105.2\pm 0.8$ m and $T=11.7\pm 0.7$ m. 

\begin{figure}
\begin{center}
\includegraphics[width=\linewidth]{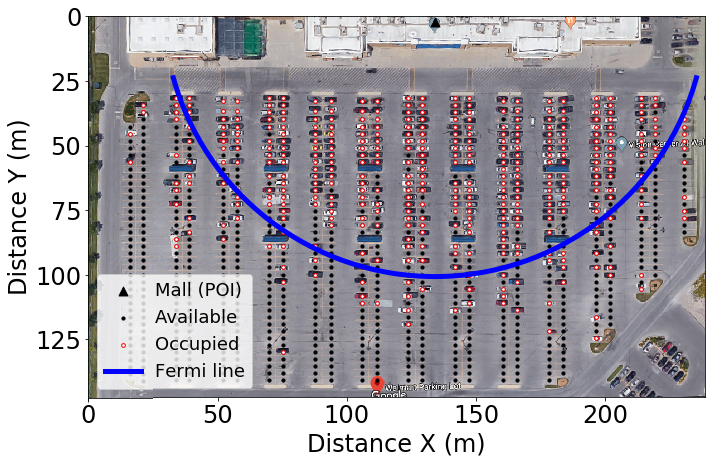}
\caption{Processed satellite photography of a mall parking lot \citep{parkimage}. Occupied parking spots are marked with red dots and empty spots in green. The Fermi line is depicted in blue.}
\label{parkphoto}
\end{center}
\end{figure}

\begin{figure}
\begin{center}
\includegraphics[width=0.9\linewidth]{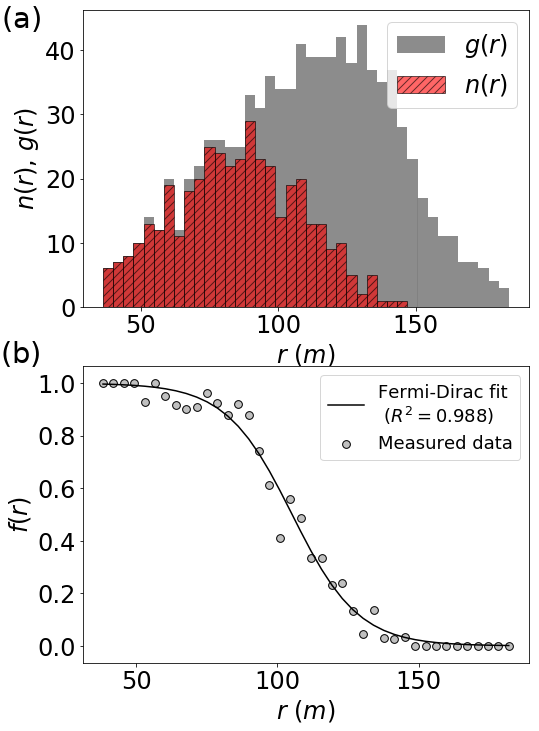}
\caption{(a) Plot of the number of occupied parking spots  $n(r)$ and of the possible states $g(r)$ per distance interval as a function of distance. The area of Figure \ref{parkphoto} has been divided into 40 radial intervals from $r_0$ to the maximum distance available. (b) Plot of the distribution function $f(r)=n(r)/g(r)$ vs. $r$. The dots represent measured data obtained from Figure \ref{park_ab}a and the solid line a non-linear regression of the data with Equation \ref{FD2}. The best fit parameters are $r_F=105.2\pm 0.8$ m and $T=11.7\pm 0.7$ m.}
\label{park_ab}
\end{center}
\end{figure}

\section{Discussion}
The proposed model is based on the above-mentioned spatial exclusion principle stemming from interactions which can be modeled in dynamical systems by means of short range repulsive forces (see, for example, \cite{zeng_specification_2017}). We will not delve into it and just recognize the resulting steric hindrance. In the case of Example I it is reflected in the mosh-pit density $\rho_m$, which will take into account effective packing plus other possible factors such as struggle for a comfortable space, visibility, etc. Example II is simpler due the fact that the cars can only park in the designed parking spots.

In both examples the fit to the Fermi-Dirac distribution (Equation \ref{FD2}) is in agreement with the measured data, so we can state that the model quantitatively describes the observed distributions to a good degree. A deeper look at the fit regression residuals expose a heavy tail in the experimental data of Example I (not shown but evident from Figure \ref{conc_ab}b). This non-random effect is not apparent in Example II. Heavy tails are commonplace in systems dealing with humans due to their intrinsic heterogeneity and inner degrees of freedom \cite{barabasi_origin_2005, wang_heavy-tailed_2017}. In this case, the deviation could be thought of being due to people which is not genuinely interested in the concert or happen to be around that add a background for large $r$. This could be taken into account by, for example, using two types of populations. However we think that the increase in the model complexity would obscure its interpretation and the insight to be gained is not worth it.

The distribution function (Equation \ref{FD2}) is defined with the parameters $r_F$ and $T$. In this simple model their meaning is easily interpretable:  $r_F$ can be seen as the radius up to which all states will be occupied and from which all will be empty in the low temperature limit $T=0$. In the concert Example I, using this fact and the calculated value for $\rho_m$ we can obtain a first estimate of $r_F$ using Equation \ref{NrF}:

\begin{equation}
r_F=\sqrt{\frac{2N}{\rho_m\alpha}+{r_0}^2}=40\pm 3\mathrm{\: m}
\end{equation}

\noindent All the audience could then be efficiently packed inside the Fermi circumference defined by this $r_F$ (see Figure \ref{concertexp}).

We considered then that the chemical potential $\mu(T)$ was constant and equal to $r_F$ at $T\neq 0$. In this simplified scenario, $r_F$ is also the distance, at any $T$, at which the occupancy is $f(r_F)=0.5$. In the case of the regression of the experimental data to the proposed distribution we obtained a value of $r_F=37.3\pm 0.3$ m, which agrees with our previous result and with the supposition that $\mu(T)$ can be considered constant (the next term in the Sommerfeld power expansion would go as $\left(T/r_F\right)^2\sim 10^{-2}$ \cite{marder_michael_p_condensed_2010}).

It is worth pointing out here that the value of the mosh-pit density $\rho_m=3.7\pm 0.7$ people/m obtained by means of the slope of $n(r)$ at $r_0$ is in very good agreement with typical values found in the literature \cite{still_introduction_2014,watson_how_2011}.

For Example II we can proceed similarly. With the positional data of the parking spots and the quantity of cars ($N=405$), we can locate the 405th closest parking spot which is at 106.6 m, so that if all cars would be parked in the most efficient way towards the POI we would have a $r_F=106.6$ m (at $T=0$). In the regression performed in Figure \ref{park_ab} we obtained $r_F=105.2 \pm 0.8$ m for the distribution at $T\neq 0$, confirming again the validity of the approximations employed.

The meaning of the temperature, while clear, is more abstract. $T$ controls the sharpness of the decay of the distribution. $T$ gives an idea of the width of this decay around $r_F$, so that $f(r)$ goes from values close to unity to values close to zero in the interval $r_F\pm T$. In practice this could give an idea of the `eagerness' of the people about the POI. For example, in Example I, very low temperatures would imply a very compact audience and therefore a very enthusiastic public. At high $T$, the public would be more spread out, pointing towards a not so devoted crowd. As in the physical counterparts, the limit of high $T$ could be well described by classical Boltzmann statistics. 

It is worth mentioning that one of the premises of the model is that $r$ has not an upper bound and deviations would appear in closed spaces if the Fermi distance $r_F$ is too close to the borders. As it can be seen in Figures \ref{concertexp} and \ref{parkphoto} where the Fermi circumferences are shown in blue, this is not the case in the examples here developed.

We want to point out that the direct application of statistical mechanics distributions to fields outside physics has had many successes. 
However it is the Boltzmann statistics the one leading the way in this regard. It is worth explicitly noting here that a simple exponential decay such as the one stemming from a classical Boltzmann distribution would clearly not describe the data in Figures \ref{conc_ab}b and \ref{park_ab}b. The application of quantum statistics is not so widespread although some other recent examples can be found related to social and economical systems \cite{rashkovskiy_bosons_2019, rosenblatt_symmetry_2019}. The type of systems described in the present work can be clearly modeled with a fermion-like statistics providing at the same time a highly illustrative and interesting application of it. To some extent, it could even be of pedagogical interest.

Summarizing, we have shown that this model provides a way to predict the distribution of people around POIs. This could be used to better design safe and comfortable urban spaces, as a starting point for the simulation of emergency situations \cite{schadschneider_evacuation_2011}, for improving the size and location of emergency exits, for the calculation of a venue of events capacity or even to measure the `temperature' of an event as described above.

\section{Conclusions}
By proposing a simple model we have determined that the spatial distribution of people around POIs closely follows fermion-like statistics in the conditions presented. The distribution is defined by two easily interpretable parameters, $r_F$ and $T$, and the model has been successfully applied to two different situations: people in an open-air concert and to cars in a mall parking lot. In both cases the agreement is evident and proves the applicability of the developed framework to real scenarios. 
\section*{References}
\bibliography{mybibfile}

\end{document}